\documentstyle[12pt]{article}
\begin{document}
\thispagestyle{empty}
\newcommand{\be}{\begin{equation}}
\newcommand{\ee}{\end{equation}}
\newcommand{\sect}[1]{\setcounter{equation}{0}\section{#1}}
\newcommand{\vs}[1]{\rule[- #1 mm]{0mm}{#1 mm}}
\newcommand{\hs}[1]{\hspace{#1mm}}
\newcommand{\mb}[1]{\hs{5}\mbox{#1}\hs{5}}
\newcommand{\bea}{\begin{eqnarray}}
\newcommand{\eea}{\end{eqnarray}}
\newcommand{\wt}[1]{\widetilde{#1}}
\newcommand{\und}[1]{\underline{#1}}
\newcommand{\ov}[1]{\overline{#1}}
\newcommand{\sm}[2]{\frac{\mbox{\footnotesize #1}\vs{-2}}
		   {\vs{-2}\mbox{\footnotesize #2}}}
\newcommand{\prt}{\partial}
\newcommand{\eps}{\epsilon}\newcommand{\p}[1]{(\ref{#1})}
\newcommand{\R}{\mbox{\rule{0.2mm}{2.8mm}\hspace{-1.5mm} R}}
\newcommand{\Z}{Z\hspace{-2mm}Z}
\newcommand{\cd}{{\cal D}}
\newcommand{\cg}{{\cal G}}
\newcommand{\ck}{{\cal K}}
\newcommand{\cw}{{\cal W}}
\newcommand{\vj}{\vec{J}}
\newcommand{\vl}{\vec{\lambda}}
\newcommand{\vz}{\vec{\sigma}}
\newcommand{\vt}{\vec{\tau}}
\newcommand{\vw}{\vec{W}}
\newcommand{\poiss}{\stackrel{\otimes}{,}}


\newcommand{\NP}[1]{Nucl.\ Phys.\ {\bf #1}}
\newcommand{\PLB}[1]{Phys.\ Lett.\ {B \bf #1}}
\newcommand{\PLA}[1]{Phys.\ Lett.\ {A \bf #1}}
\newcommand{\NC}[1]{Nuovo Cimento {\bf #1}}
\newcommand{\CMP}[1]{Commun.\ Math.\ Phys.\ {\bf #1}}
\newcommand{\PR}[1]{Phys.\ Rev.\ {\bf #1}}
\newcommand{\PRL}[1]{Phys.\ Rev.\ Lett.\ {\bf #1}}
\newcommand{\MPL}[1]{Mod.\ Phys.\ Lett.\ {\bf #1}}
\newcommand{\BLMS}[1]{Bull.\ London Math.\ Soc.\ {\bf #1}}
\newcommand{\IJMP}[1]{Int.\ J.\ Mod.\ Phys.\ {\bf #1}}
\newcommand{\JMP}[1]{Jour.\ Math.\ Phys.\ {\bf #1}}
\newcommand{\LMP}[1]{Lett.\ Math.\ Phys.\ {\bf #1}}

\renewcommand{\thefootnote}{\fnsymbol{footnote}}
\newpage
\setcounter{page}{0}
\pagestyle{empty}
\vs{12}
\begin{center}
{\LARGE {\bf Super-Affine Hierarchies}}\\
\vs{6}
{\LARGE{\bf and their Poisson Embeddings}}
\\[0.8cm]

\vs{10}
{\large Francesco Toppan}
\quad \\
{\em (UFES, CCE Depto de F\'{\i}sica,
Goiabeiras cep 99060-900, Vit\'{o}ria (ES), Brasil)} 
\quad \\
{\em Talk given in Dubna, July 1997, in memory of V.I. Ogievetsky}
\end{center}
\vs{6}

\centerline{\large {\bf Abstract}}
\vs{3}
\noindent

The link between (super)-affine Lie algebras as Poisson brackets 
structures and integrable hierarchies provides both a classification 
and a tool for obtaining superintegrable hierarchies.
The lack of a fully systematic procedure for constructing
matrix-type Lax operators, which makes the supersymmetric case
essentially different from the bosonic counterpart, is overcome via
the notion of Poisson embeddings (P.E.), i.e. Poisson mappings 
relating affine structures to conformal structures (in their simplest 
version P.E. coincide with the Sugawara construction).
A full class of hierarchies can be recovered by using uniquely
Lie-algebraic notions.
The group-algebraic properties implicit in the super-affine picture 
allow a systematic derivation of reduced hierarchies by imposing 
either coset conditions or hamiltonian constraints (or possibly both).
 
\vs{6}
\vfill
\rightline{DF/UFES-P002/98}
{\em E-Mails:\\ toppan@cce.ufes.br, toppan@cat.cbpf.br}
\newpage
\pagestyle{plain}
\renewcommand{\thefootnote}{\arabic{footnote}}
\setcounter{footnote}{0}

\noindent{\section{Introduction.}}

\par
Affine Lie algebras and conformal algebras have received a great 
attention
in the physicists community in the last several years, mainly due 
to their
relevance to phenomenological models ($2d$ $\sigma$-models in 
the WZNW
description), as well as the more fundamental string approach to the
unification of interactions.
It is well understood by now that conformal algebras (even the 
non-linear
$W$-type ones) are the output of affine algebras after 
some construction,
hamiltonian reductions or cosets, are performed on them.
\par
While affine-Lie and conformal algebras are universally appreciated, 
not so
much attention has received a truly remarkable property they share, 
i.e.
that they support hierarchies of integrable equations in 
$1+1$ dimension
in the sense that they provide (one of the) Poisson Brackets 
(PB for short in
the following) for the associated hierarchy.\par
To my knowledge such a property has not yet found a direct 
implementation
in the string-theory program, however has already found application to
physically motivated theories like discretized $2d$ gravity in the
matrix-model approach (see \cite{dgz} and references therein).
There, essentially,
$W$-algebras arise as Ward identities known as $W$-constraints and the
partition function is a tau-function of an associated integrable 
hierarchy.\par
Moreover integrable hierarchies underline such exactly solvable 
models as
$4d$ $N=2$ Seiberg-Witten SYM theories (see e.g. \cite{sw}).\par
Due to the above-mentioned results it is clear why a lot of 
attention
continues to be focused on the supersymmetric extensions. 
It is hoped that
their understanding will provide the basis for discretized 
$2d$ supergravity
(see \cite{alv}). However, unlike the purely bosonic theories, the
supersymmetric extensions, for reasons we will discuss later, 
have so far
failed being accomodated into a single unifying picture.
Due to this basic problem super-hierarchies have been produced by 
using all
sort of tools available, i.e. by direct construction,
via Lax operators, bosonic as well as fermionic
and
both in scalar or matrix form, 
by coset procedure
\cite{{man},{mat}, {IK}, {BD},{top},{del}, {bon}} and so on.
We have by now an impressive list of ``zoological'' data concerning
superhierarchies. There is an overwhelming evidence that some order 
should be
made and a single unifying picture should be provided.
In this paper we wish to point out a possible tool for both
classifying and explicitly constructing a class of super-hierarchies. 
\par
Such a tool is based on the super-affine framework
and Poisson Embedding (or shortly PE). This means
the derivation of super-hierarchies by regarding as fundamental
ingredient the Poisson brackets structure furnished by the
supersymmetric affinization of a (super)-Lie algebra (more on this
notion later). Poisson Embeddings are a special class of Poisson 
mappings,
(i.e. maps between two sets of (super)-fields 
$f_P: \{\Phi_{i}\}_{I} \mapsto
\{{\tilde{\Phi}}_j\}_{II}$ which preserves the PB structures 
between sets
$\{I\}$ and $\{II\}$) having the further property that the PB 
structure in
$\{II\}$ is the one of a given integrable hierarchy.\par
As a consequence it is possible to define on the super-fields in 
$\{I\}$
a hierarchy of equations which inherit the integrability property
from the one defined on the second set $\{II\}$.
This seemingly innocent remark has indeed
far-reaching consequences and allows us to produce and identify new 
hierarchies
of equations, even generalizing the set of hierarchies produced in the
literature with more dispendious and time consuming methods.\par
Well-known examples of Poisson maps are the Wakimoto free 
(super)-fields
realization of affine algebras, and the 
Sugawara-type construction relating
affine(super-)fields to the (super)-stress energy tensor. 
The latter is also
a Poisson Embedding, and therefore integrable hierarchies are induced
both at the level of affine and of Wakimoto free superfields. \par
A real breakthrough in this context appears to be the realization 
in \cite{IKT}
that among such mappings there is a (differential) polynomial 
Poisson
map which is an $N=4$ extension of the Sugawara construction based on
super-affine
$sl(2)\oplus u(1)$. Such a mapping, besides being a PB one, 
is a Poisson
Embedding since the Sugawara-produced hierarchy turns out to be the 
small
$N=4$ SCA carrying the $N=4$ KdV hierarchy.\par
The key observation is that superconformal algebras can be more 
directly 
identified with the PB structure of a given hierarchy since they are
easily accomodated into scalar-type Lax operators (see \cite{del}), 
which can be
constructed with a systematic procedure.\par
Focus can therefore be put into the construction of generalized
Sugawara mappings.
Here however we advocate the point of view that we
can investigate the properties of {\em already existing} Sugawara
constructions to identify and classify series of new hierarchies.
In particular one of such mappings sends any 
$N=2$ (super)-affine algebra
into $N=2$ Virasoro. This is the PB structure for three distinct 
$N=2$ KdV
hierarchies \cite{mat} (associated to a value of parameter
$a=4,-2,1$). Accordingly, three induced hierarchies are associated to 
the
$N=2$ affine superfields realizing the PE. In some cases, 
the three hierarchies collapse into a single one.\par
The full power of affine algebras gets really appreciated when one 
realizes
that due to stringent group-theoretical reasons one can further 
reduce such
algebras with coset procedures and/or hamiltonian reductions. 
On conformal
algebras themselves these procedures cannot be carried out. For 
instance
the Virasoro algebra itself is already a hamiltonian 
reduction of affine $sl(2)$ \cite{pol}.
\par
A full bunch of ``popping out'' hierarchies can therefore 
find a natural
group-theoretical explanation and interpretation. 

\vspace{0.2cm}
\noindent
{\section{Notations and preliminary remarks.}}

\par
The class of Poisson brackets structure we will consider is given 
by the
superaffinization of any given semisimple (super)-Lie algebra, 
defined
as follows:
Let ${\cal G}$ be any finite semisimple Lie algebra, 
either purely bosonic or
supersymmetric, with $n_b$ bosonic and $n_f$ 
fermionic generators ($n_f=0$
for standard Lie algebras) collectively denoted as $\tau_{\alpha}$,
for $\alpha= 1, ..., n_b+n_f$, and let 
${f^\gamma}_{\alpha \beta}$ denote
the structure constants.\par
We can introduce the $N=1$ superfields (in the
superspace coordinate $X= x, \theta$, with $\theta$ Grassmann)
$\Psi_\alpha (X)$, associated to each generator 
$\tau_\alpha$ and with opposite
statistics w.r.t. $\tau_\alpha$.\par
The superaffine algebra is defined by assuming 
the following Poisson brackets
\begin{eqnarray}
\{ \Psi (X)_\alpha, \Psi (Y)_\beta\} 
&=_{def}&{ f^\gamma}_{\alpha\beta}
\Psi (Y)_\gamma \delta (X,Y) + c K_{\alpha\beta}D_Y\delta(X,Y)
\label{pb}
\end{eqnarray}
where we introduced the supersymmetric Dirac's $\delta$-function
\begin{eqnarray}
\delta (X,Y) &=&\delta(x-y)(\theta -\eta)\nonumber
\end{eqnarray}
and the $N=1$ superderivative
\begin{eqnarray}
D_Y &=& {\textstyle{\partial\over\partial \eta}}+
{\eta}{\textstyle{\partial
\over\partial y}}
\nonumber
\end{eqnarray}
for $Y\equiv y, \eta$.\par
$c$ is the central extension and $K_{\alpha\beta}$ is 
defined as a supertrace
$K_{\alpha\beta}= Str({\tau_\alpha}\tau_{\beta})$ 
in a given representation
for ${\cal G}$, let's say the adjoint.\par
In the above relation the Jacobi identities are satisfied and 
therefore
(\ref{pb}) indeed defines a PB structure.\par
We mention that if the (super)-algebra ${\cal G}$ 
of departure admits a
complex structure, then (\ref{pb}) is 
indeed $N=2$ supersymmetric and can be
recasted into a manifestly $N=2$ formalism
(see \cite{iva}, here however we do not
need such technical improvement.\par
The bosonic limit of the above (\ref{pb}) affine algebra 
(realized on a
bosonic ${\cal G}$ and via purely bosonic fields) is the 
building block
for constructing generalized Drinfeld-Sokolov hierarchies, via the
association to a matrix type Lax operator $L$ of the kind
\begin{eqnarray}
L &=& \partial +\sum_{\alpha} J_\alpha (x) \tau_\alpha + \Lambda
\end{eqnarray}
where $\Lambda$ is a constant element in the
${\tilde {\cal G}}={\cal G} \otimes {\bf C}(\lambda, 
\lambda^{-1})$ loop
algebra of ${\cal G}$ realized on an auxiliary variable 
$\lambda$ which
plays the role of a spectral parameter. By DS construction, 
if $\Lambda$
is chosen in such a way to realize the decomposition
\begin{eqnarray}
{\tilde{\cal G}} &=& {\cal K}\oplus {\cal M}
\label{split}
\end{eqnarray}
with ${\cal K} = Ker_{ad-\Lambda}$ and ${\cal M}= Im_{ad-\Lambda}$
over the adjoint action of $\Lambda$, and if furthermore ${\cal K}$
is abelian
\begin{eqnarray}
\relax [ {\cal K}, {\cal K}] = 0
\label{abel}
\end{eqnarray}
then we are guaranteed about the existence of infinite integrals of 
motion
in involution.\par
The same kind of construction has been generalized in 
the supersymmetric case
by Inami and Kanno in a series of papers 
(see e.g. \cite{IK} and references
therein). Now $L$ assumes the form
\begin{eqnarray}
L &=& D + \sum_{\alpha} \Psi_{\alpha} (X) \tau_\alpha + \Lambda
\label{superlax}
\end{eqnarray}
the first and second terms in the r.h.s. are fermionic and so 
$\Lambda$
must be fermionic as well. This constraint puts a very strong 
restriction
on the
superhierarchies which can be obtained through DS procedure. 
In the bosonic
case for instance generalized-KdV hierarchies which include 
among others
Boussinesq
are defined by taking $\Lambda$ to be the sum over all simple roots
of the ${\tilde{\cal G}}$ algebra\footnote{this is a slightly
imprecise way of saying, in order to talk about simple roots 
we should
introduce the extended affine algebra over the
auxiliary loop space parameter $\lambda$, 
but let's avoid these technical
complications here.}. Consequently generalized 
super-KdV hierarchies can be
obtained solely from those superalgebras which admit a Dynkin diagram
presentation
involving only fermionic simple roots. Admittedly, this is a rather
restrict class of superalgebras.\par
For them however we have a viable and systematic 
procedure to construct
superhierarchies.\par
A problem arises because very natural hierarchies like the 
supersymmetric
extensions of NLS fail to be accomodated into this scheme due to the 
fact
that their bosonic equivalents are recovered from a $\Lambda$ in
(\ref{superlax}) belonging to
the Cartan generators of ${\cal G}$. Since the Cartan sector of 
whatever
simple bosonic Lie or super-Lie algebra is in any case bosonic, 
we have no
possibility at all to construct a fermionic $\Lambda$ with the desired
properties.\par
In effect there exists a class of superalgebras, the so-called strange
superalgebras of the $Q(n)$ series, which admit a fermionic
Cartan sector (see \cite{sor}). But these superalgebras are of no use
here for another reason. 
The fermionic $\Lambda$ taking value in the fermionic
Cartan always fail to satisfy either condition (\ref{split}) or 
condition
(\ref{abel}).\par
There are some {\em ad hoc} procedures to overcome this difficulty
(see {\cite{top2}), but it must be said that even if viable for the 
practical
purpose of computing higher order hamiltonians, they lack a clear and
compelling motivation which makes them not attractive for the
purpose of classifying hierarchies. Similarly, 
matrix-type Lax operators
have been produced for hierarchies of super-NLS type (or 
obtainable from
super-NLS) reduction. 
Such Lax operators, unlike the bosonic ones, have
entries which are composite superfields. 
Here again it is hard to justify
their appearance in terms of fundamental principles. Rather, they are
the signal that a simpler structure should be found behind them.
In the next section we will see how an answer (at least a partial one)
can be provided.

\vspace{0.2cm}
\noindent{\section{Poisson maps and Poisson embeddings.}}

\par
In the previous section we have discussed some problems arising 
in the
construction of superintegrable hierarchies. Now we will show how to
overcome such problems via the introduction of 
the notion of Poisson maps
and Poisson embeddings, already outlined in the introduction. 
Evidence will
be furnished that superaffine algebras are the right setting 
to deal and
classify superhierarchies.\par
A point should be clear, the maps we are investigating are 
{\em polynomial
differential} maps. In literature non-polynomial maps 
relating different hierarchies have been considered
(in some cases they are not even Poisson maps),
but in all known examples they can be recasted into
(or derived from) polynomial differential
maps. So it seems there is no compelling reason to look 
beyond the realm
of polynomial differential Poisson maps.\par
Under an $f_P$ Poisson Embedding the infinite series of 
$H_k$ hamiltonians in
involution
of the integrable hierarchy ($\{H_k, H_{k'}\}_{(II)}=0 $) can 
be regarded
as hamiltonians in involution w.r.t. the first PB structure
(for any couple is indeed $\{H_k, H_{k'}\}_{(I)} =0$) and it 
makes sense to
define
an infinite series of compatible flows for the superfields $\Phi_i$ 
as:
\begin{eqnarray}
{\textstyle{\partial\over\partial t_k}}\Phi_i &=& \{ H_k
(f_P(\Phi_j)), \Phi_i\}_{(I)}
\end{eqnarray}
We will refer to hierarchies of this kind either as the induced or as 
the
embedded hierarchy.\par
Examples of PE are given by Sugawara-type construction. We recall 
that
for any bosonic and $N=1$ supersymmetric affine algebra there 
exists a 
well-defined procedure which allows us to produce conformal fields
satisfying a closed
${\cal W}$-algebra structure. They are expressed in terms of the
enveloping algebra of
the (super)-affine algebra ${\cal G}$ and are in $1$-to-$1$
correspondence with each Casimir of ${\cal G}$. The most relevant or 
leading term in the
enveloping algebra being given by
\begin{eqnarray}
&& d^{i_1...i_n}J_{i_1}(x)...J_{i_n}(x)\nonumber
\end{eqnarray}
in the bosonic case and
\begin{eqnarray}
&& d^{i_1...i_n}(D_X\Psi_{i_1})\cdot ...\cdot (D_X\Psi_{i_{n-1}})\cdot
\Psi_{i_n}(X)
\end{eqnarray}
in the $N=1$ super-case. \par
Here $d^{i_1...i_n}$ is the symmetric tensor denoting an $n$-th
order Casimir and $J_{i_k}$ ($\Psi_{i_k}$) are spin $1$ fields (spin
${\textstyle{1\over 2}}$ superfields respectively).
\par
A full bunch of improvements or
covariantization terms must be added
to the above ``leading order terms''
to render the (super)-field a primary
or conformal field. For instance, in the case of the order $2$ Casimir
(which exists for any (super)-Lie algebra)
the terms to be added are just the Feigin-Fuchs
terms which provide a non-vanishing central charge so that the
Sugawara-constructed field satisfies the full Virasoro 
($N=1$ superVirasoro)
algebra.\par
As an example, in the case of the $sl(n)$ series the $W$-algebra 
so produced
is the $W_n$
algebra (or its $N=1$ supersymmetrization). The ``miracle''
here, which has a group-theoretical explanation, lies in the fact 
that this is the same algebra arising from Dirac brackets after the
Drinfeld-Sokolov hamiltonian reduction sketched in the previous 
section has
been taken into account.\par
For the moment let me just point out that the (super)-affine 
algebra, which
at the beginning was not associated to any evident hierarchy, 
has now acquired
the status of a PB structure for the induced hierarchy. Any 
Poisson map onto
the super-affine fields is therefore also a Poisson Embedding.\par
There exists a well-defined prescription on how to realize any
super-affine algebra in terms of free (super)-fields.
The result is given by the
(generalized)
(super)-Wakimoto realizations whose origin is traced on the 
theory of
non-linear realizations of algebraic structures.
The simplest case of a super-Wakimoto construction is the 
realization of
the
$N=1$ affine $sl(2)$ algebra \cite{top}.\par
The generalized (super)-Wakimoto realizations are all 
examples of Poisson maps.
In full generality the Wakimoto free (super)-fields satisfy induced
hierarchies whose origin arises from their ``double'' 
embedding into the
generalized KdV hierarchies.\par

\vspace{0.2cm}
\noindent{\section{Induced $N=2$ and $N=4$ hierarchies.}}

\par
In the previous section we have discussed the general theory of 
Poisson
Embeddings and have shown that super-affine Lie algebras are 
associated
with (at least $N=1$) integrable hierarchies.\par
Here we further discuss properties of PE and study them in the 
context of
$N=2$ hierarchies. This is indeed a very interesting case 
since for the
first time appears 
(as shown by explicit construction in \cite{mat})
that one and the same PB structure is associated with different 
series
of integrable hierarchies.
\par
Indeed if we denote as $J(x,\theta , {\overline \theta})$ 
the real $N=2$
Virasoro field consisting of two (spin $=1,2$) boson 
components and a couple
of spin ${\textstyle{3\over 2}}$ fermions, three inequivalent $N=2$
KdV hierarchies
can be recovered (for a parameter $a$ taking
values $a = 4, -2, 1$ respectively).\par
Therefore any Poisson map onto $N=2$ Virasoro induces three 
different and in
principle inequivalent series of hierarchies.\par
The case $N=2$ is interesting also for another reason. While 
generalized $N=1$
Sugawara poses no problem in its construction (and in its 
association with
super-KdV hierarchies), much less is known concerning the $N=2$
case. Apart from the Sugawara construction of the $N=2$ Virasoro
discussed below, to my knowledge no general theorem has 
been given so far and
only explicit examples have been worked out on how to perform 
Sugawara
construction of 
$N=2$ ${\cal W}$-algebras (the main question concerns
the feasibility of adding Feigin-Fuchs terms, while mantaining 
a full $N=2$
${\cal W}$ algebra structure).
\par
Neverthless what we already have 
can be exploited
to produce and
identify interesting classes of induced $N=2$ hierarchies.\par
Let us for the moment discuss just the simplest examples of
superaffine algebras producing PE onto $N=2$ Virasoro. They are given
respectively by the $N=1$ affinization
(the following 
examples are
$N=2$ supersymmetric and can be reformulated with an
$N=2$ formalism as stated before) of
\par
{\em i)} the $u(1)\oplus u(1)$ algebra;\par
{\em ii)} the $sl(2)\oplus u(1)$ algebra and\par
{\em iii)} the $sl(2|1)$ superalgebra.
\par
In the first two examples the original algebra is not a 
simple one,
this is just because we need a complex structure (provided by the 
extra $u(1)$)
which enables us to have a second supersymmetry.\par
$sl(2|1)$ is the simplest example of an $N=2$ Lie superalgebra; it 
contains
$sl(2)$ as subalgebra and four extra fermionic simple roots.
\par
The first case coincides with the well-known $N=2$ version of 
(the three induced hierarchies of) m-KdV.\par
Much more interesting and rich of structure is case {\em ii)}. 
The algebra
here admits a quaternionic structure which allows a full $N=4$
Sugawara construction
(\cite {IKT}). This has the consequence that,
besides the induced $N=2$ hierarchies (for $a= 4,2,-1$) recovered
from the $N=2$ Sugawara through, respectively,
{\em a)} the $N=2$ Virasoro $J_{u(1)\oplus u(1)}$ associated to the
$u(1)\oplus u(1)$ subalgebra (it coincides with the 
previous case),
{\em b)} the full $N=2$ Virasoro $J_{full}$ given by\\
$
J_{full} = H{\overline H} + F {\overline F}
+ c D {\overline H} + {\overline c }
{\overline  D} H
$\\
(where the last two are the Feigin Fuchs terms),
{\em c)} the coset $N=2$ Virasoro given by\\
$
J_{coset} =J_{full}-J_{u(1)\oplus u(1)}=  
F {\overline F} + c' D{\overline H}
+ {\overline c}' {\overline D} H,
$\\
we can defined induced an induced hierarchy based on $N=4$
KdV. This hierarchy has a very interesting property, namely 
that it is consistent with the equations of motion to set
\begin{eqnarray}
 && H={\overline H}=0.
\end{eqnarray}
The reduced hierarchy on the superfields $F, {\overline F}$
coincides with the $N=2$ NLS.
\par
In this framework the NLS hierarchy can therefore be directly obtained 
from its Poisson Embedding properties on $N=4$ KdV, in contrast
for instance with
the original coset construction  
of $N=2$ NLS \cite{top}, whose superfields were mapped onto an $N=2$ 
Virasoro {\it without} central charge and for 
that reason integrability had to be proven separately and with 
different methods.\par
The {\em iii)} case can be treated similarly. This superalgebra 
contains
$sl(2)\oplus u(1)$ as a subalgebra. All the induced hierarchies 
defined
in the previous case can therefore be consistently extended. 
They involve extra equations of motions 
associated
to the $N=2$ non-linearly (anti)-chiral bosonic superfields of
spin ${\textstyle {1\over 2}}$ (of ``wrong'' statistics)
associated to the fermionic roots.\par 
It is clear at this point that a full class of
$N=4$ induced hierarchies can be produced from the superaffinization 
of any quaternionic super-Lie algebra ${\cal G}_Q$
(i.e. with $N=4$ supersymmetric PB),
with the following recipe:\par
{\em i)} individuate any $sl(2)\oplus u(1)$ subalgebra,\par
{\em ii)} construct from the given subalgebra the $N=4$ 
Sugawara 
leading to $N=4$ SCA,\par
{\em iii)} use this concrete realization of the $N=4$ SCA as 
PB structure
for the $N=4$ KdV.\\
An $N=4$ hierarchy is automatically induced on the affine superfields 
generating the full ${\cal G}_Q$ algebra. The induced hierarchy
is automatically $N=4$ invariant because by construction {\em both} 
the hamiltonians and the superaffine-${\cal G}_Q$ PB structure 
are $N=4$ supersymmetric.\par
Furthermore, on these induced hierarchies it is possible to 
investigate
whether consistent reductions can be imposed both as hamiltonian
constraints or coset construction.
\vspace{0.2cm}
\noindent{\section{Conclusions.}}

In this talk I have investigated the role of Lie-algebraic methods
in classifying integrable hierarchies. One of the nice features
of the approach based on super-affine Lie algebras is that it
allows investigating systematically hamiltonian constraints and
coset reductions. The structure of $N=4$ integrable hierarchies is
currently under investigation, with the methods here outlined,
in a collaboration with E. Ivanov and S. Krivonos. 

\vspace{0.2cm}
\par
\noindent{\large{\bf{Acknowledgments}}}

\par
I express my gratitude to the organizers of the workshop in memory
of V.I. Ogievetsky for their invitation.\par
This work has been written while the author was under a JSPS 
(Japan Society for the
Promotion of Science) contract. I am very grateful to the members
of the Physics Dept. at Shizuoka University for their warm and kind
hospitality. I am especially pleased to thank Shogo Aoyama for his 
much needed constant help and support.  Finally, I have profited of 
clarifying discussion with E. Ivanov and S. Krivonos.

\vspace{0.2cm}

\section{\bf Appendix: the classification of cosets.}

\par
One of the nice features of the approach based on affine superfields
is the fact discussed in the text that it allows to construct
reduced hierarchies by exploiting symmetries and group-theoretical
properties of the affine algebras. Here we will discuss
the class of reductions known as ``cosets'', which arises when 
superfields
associated to semisimple (super)-Lie 
subalgebras are set equal to zero.
This class of reductions are classified by all inequivalent
(super)-Lie subalgebras which can be embedded into ${\cal G}$, the 
(super)-Lie
algebra whose superaffinization furnishes the original hierarchy.\par
The classification scheme
to find all inequivalent Lie subalgebra of a given Lie algebra has 
been given
by Dynkin. This scheme provides all possible cosets we 
can construct out
of a given hierarchy. To be specific
we specialize our discussion here to the $sl(3)$ case which already 
contains
all the features we are interested in.\par
The full list of $sl(3)$ subalgebras is given by
\begin{eqnarray}
&& u(1),\quad u(1)\oplus u(1),\quad
sl(2),\quad sl(2)\oplus u(1)
\end{eqnarray}
The corresponding coset hierarchies involve $8-1=7$, $8-2=6$, $8-3=5$
and $8-4=4$ (super)-fields respectively ($8$ is the order of $sl(3)$).
This is not the full story however because an abelian $u(1)$ 
generator
($=v$) can always be chosen to lie in the Cartan sector of $sl(3)$ 
and be
specified by an angle $\phi$:
\begin{eqnarray}
v &=& cos\phi H_1 + sin\phi H_2,
\end{eqnarray}
where $H_1,H_2$ are the
two Cartan generators of $sl(3)$. \par
Coset hierarchies w.r.t. an abelian $u(1)$
are therefore labelled by such an angle $\phi$. The angle however is 
not
completely arbitrary and is further required to lie on the interval
\begin{eqnarray}
&& 0\leq \phi < {\textstyle {\pi\over 6}}
\end{eqnarray}
The reason is the presence of the
extra discrete symmetries which involve the Cartan
decomposition of a Lie algebra. In case of $sl(3)$ there are two
such sources of symmetries:\par
{\em i)}
the $Out = {\textstyle {Aut\over Int}}$
automorphism group, which coincides here with ${\bf Z}_2$ and is
related to the symmetry of the Dynkin diagram
(i.e. the exchange of the two simple roots);\par
{\em ii)} the Weyl group of $sl(3)$,
which is the $6$-elements $S_3$ permutation
group.\par
The full group of discrete automorphisms coincides here 
with the $12$-elements
group, direct product of $S_3$ and ${\bf Z}_2$. It can be easily 
realized
that its action on the Cartan subspace spanned by the $H_1$, $H_2$
generator is that of the
finite rotation group generated by $30$-degrees rotations, which
finally leads to the above restrictions on the angle $\phi $.\par
For what concerns $sl(2)$ subalgebras there exists only two 
inequivalent
ways of embedding $sl(2)$ on $sl(3)$, up to 
the $Adj$-action
of $sl(3)$ internal automorphism group.
They correspond to the two  decompositions of the $8$ $sl(3)$ 
generators in terms
of the $sl(2)$ representations.\par
Only the latter $sl(2)$ allows to accomodate a further abelian $u(1)$
subalgebra, and therefore the $sl(3)$ coset over $sl(2)\oplus u(1)$ 
is unique.\par
In this way we have listed the full class of coset-hierarchies 
arising
from $sl(3)$. The case of the coset over $sl(2)\oplus u(1)$ is 
of particular
interest because both $sl(3)$ and $sl(2)\oplus u(1)$ 
are quaternionic algebras.
It is likely that its supersymmetric extension
(under the appropriate $N=4$-invariant hamiltonians)
would correspond to an $N=4$
coset hierarchy. This case is currently under investigation in a 
collaboration with Ivanov and Krivonos .\par
The above procedure can be carried in full generality and cosets 
can
be classified according to the Dynkin scheme.\par
It is worth to notice
that no
restriction involving e.g. symmetric space is required to define 
consistent
coset hierarchies, they are just (a more specialized) example of
coset-construction.

\end{document}